# Exploring Evolving Plants as Interacting Particles in a Randomly Generated Heterogeneous Environment

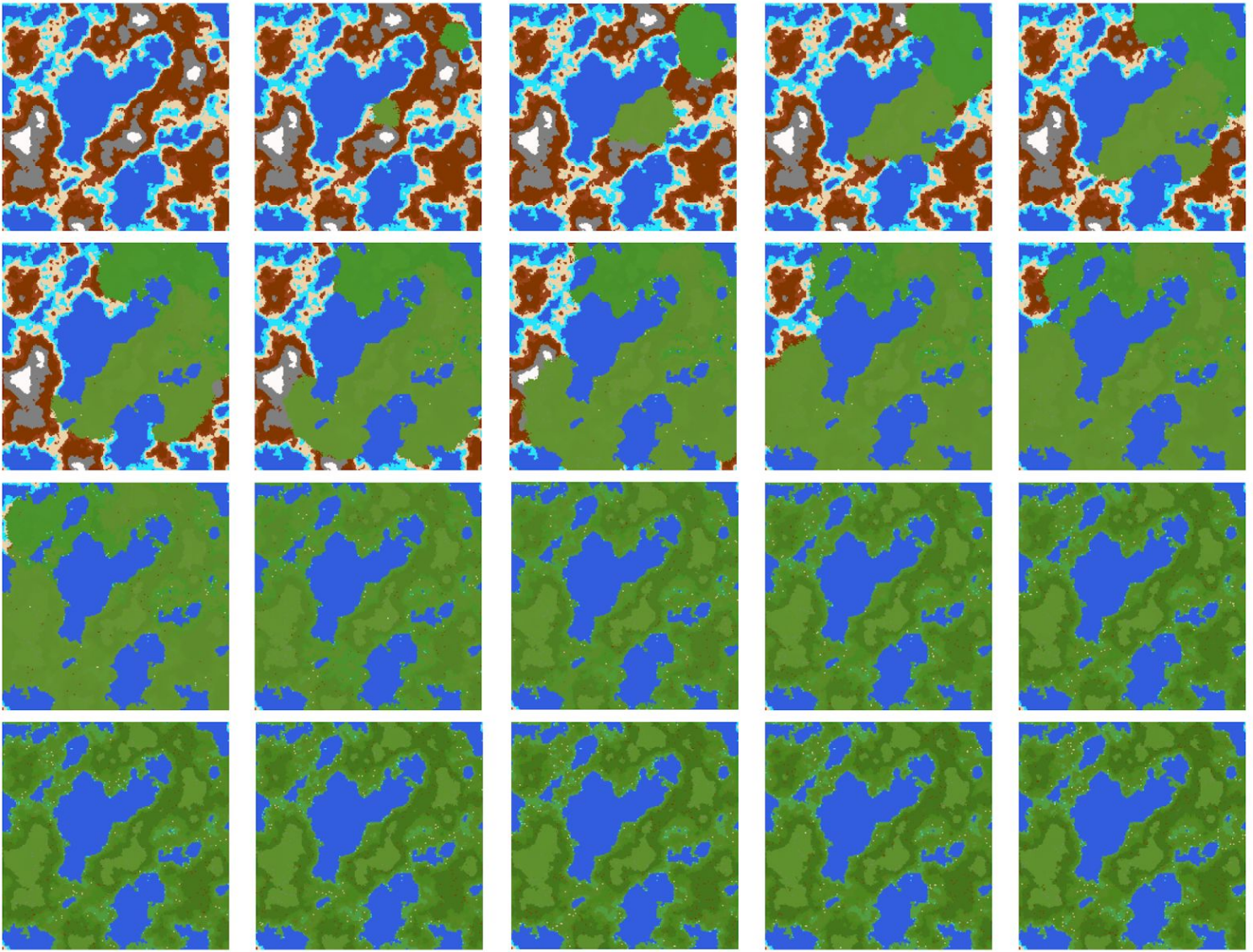

Sasha Khazatsky, Albert Yu, Zihao Zhao, and Gabe Zuckerman

## Abstract
We model evolution of plants in a world, made up of different locations, with multiple environments (mutually exclusive and collectively exhaustive subsets of locations). Each environment (landmass) has temperature, rainfall, and other attributes that directly affect plant growth and reproduction. Each plant has preferences for environment attributes. Depending on how suitable the environment is to the plants, seeds are released or death occurs. With every reproductive cycle, genetic mutations occur. To model competition, plants in compete for survival, and success is stochastically dependent on environmental fitness. Our model determines whether and how evolution occurs, and how the attributes of plants change and possibly converge over time in relation to the attributes of the environment.

## Introduction
Since Charles Darwin's *Voyage of the Beagle* (originally published in 1839) and subsequent publication *On the Origin of Species* in 1859, much time has been dedicated to the study of evolutionary theory.  Darwin claimed that species arose through natural selection, a process in which small inherited variations increase fitness--the ability to survive and reproduce--and eventually lead to speciation (Darwin, 1859).  In our simulation, we leveraged a probabilistic model to show this along with other relevant ecological theories.

We begin the construction of our random world using the Perlin noise algorithm (Perlin, 1985) for smooth land-water contours.  Our gridded world contains 7 cell types, or environments.  Listed in order of elevation, they are: blue ocean, light blue wetlands, tan mangroves, brown dirt, grassy land, gray mountains, and white snow on several islands.  We continue by creating 5 plant archetypes: mangrove, palm tree, oak tree, pine tree, and shrub.  Each plant archetype has its own preferred environment.  We use fitness vectors to denote a plant's preference for an environment.  A plant's trait vector is comprised of its fitness (or likelihood of survival and reproduction) in each of the 7 environments.

A plant can (but, does not always) reproduce once per timestep of the simulation, and produce a seed (or more than one, see *Experiments* below) in a random nearby cell.  A seed's fitness vector is similar to its parent's; however, there is a small amount of gaussian noise added to represent mutations--the heritable variations in the gene pool that drive evolution.  A seed can then grow into a plant that is able to reproduce.  If multiple plants, seeds, or a combination of the two arrive at the same gridcell, only one will survive in each timestep.  This competition is decided by a probabilistic function of the plants and seeds fitness vectors.  In each timestep, whether a plant or seed survives and reproduces is determined by respective probabilistic functions of the fitness vector.

We found that plants tended to converge to match the attributes of their environment, even after multiple disturbances. In cases where plants were not allowed to mutate, those unfit for their current environment eventually spread to areas they were more fit for. We also experimented with offspring r versus K reproduction survival strategies. Finally, we experimented with how changing the rate of plant mutations over time impacted their ability to converge to both unchanging and changing environment attributes.

## Experiments
### I. Natural Selection
To test Darwin's theory of natural selection, we begin our simulation by initializing a random number of plants at random locations, each with varying fitness vectors drawn from some normal distribution.  We test to see if the number of plants converges to the number of environment in our world. Additionally, we test if plants migrate to their ideal habitats from their randomly initialized locations without evolving.

### II. r/K Selection
Introduced by E.O. Wilson and Robert MacArthur in their 1967 publication *The Theory of Island Biogeography*, this book contained many important contributions to the field of ecology.  This theory examines the tradeoff between quantity and quality of offspring.  r-selected species produce many offspring each with low chances of survival, while K-selected species produce fewer offspring with higher likelihood of survival.  To understand how this tradeoff effects our system, we will compare 2 classes of simulations.
- Class 1: plant reproduction is limited to one offspring with a slightly more favorable fitness than its parent
- Class 2: plant reproduction is very frequent, but seeds are of varying levels of fitness

We will see how convergence differs in the two classes of simulations.

### III. Disturbance Resilience
More relevant today than ever, disturbance ecology seeks to understand how communities of organisms react to events that wipe out a significant proportion of the population.  We will see how this affects our system by adding an event that occurs with low probability, in which a high proportion of plants is killed, and see how convergence is affected.

### IV. Baseline
We have additionally created a graph-based simulation and ran appropriate simulation to address the baseline requirements.  This is attached in the appendix.

## Methods
### Our Simulation
For each simulation we randomly generate richly diverse, uninhabited environments for life to start in. As the world starts progressing, life develops stochastically throughout the environments. With the spread of plant species, mutations, modeled by gaussian noise, occur in genes. For better or worse, these gene mutations change underlying environmental fitness distributions in new plants. Survival difficulty, land distribution, island development, natural disaster frequency, world dimension, initial plant population, mutation rate, and plant reproduction type are tunable parameters for experimentation.

At each time step:
- Plants age, dying with a probability based off their respective environmental fitness
- Surviving trees release seeds according to their environmental adaptation.
- Seeds are then spread by the environment's wind and ocean currents, which are modeled with multivariate gaussians.
- Life growing in overlapping soil compete for survival in a stochastic fitness competition, and the victor is sampled from a softmax distribution of fitness levels.

**Probabilistic Functions**
**Fitness Function**

$$F = e^{-0.25(unfitness)}$$

$$unfitness = ||plantTraits - soilTraits||_2$$

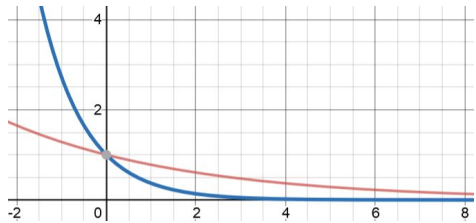

As our fitness function, we chose e^-0.25x (red) instead of e^-x (blue) because it decayed slower so that somewhat unfit plants (who had a higher x value) would have a fitness score substantially above 0. This would be useful as inputs for our competitor function.

**Competitor Function**
In our grid implementation, the probability that a seed with traits x_k would be the next tree occupant of the soil instance was given by the function, for some constant c:

$$\{S\} = \text{seeds at location (i, j)}$$

$$p = \frac{e^{c \cdot (F(x_k))}}{\Sigma_{i \in \{S\}} e^{c \cdot (F(x_i))}}$$

Occasionally, this led to exploding exponentials, so we modified the formula to a numerically equivalent equation:

$$\{S\} = \text{seeds at location (i, j)}$$

$$m = \max_{i \in \{S\}} F(x_i)$$

$$p = \frac{e^{c \cdot (F(x_k)) - m}}{\Sigma_{i \in \{S\}} e^{c \cdot (F(x_i)) - m}}$$

In our graph implementation, this probability was calculated with a different function, which gave unfit plants a larger chance of being selected. This helped ensure that our convergence model was robust.

$$\{S\} = \text{seeds at node k}$$

$$p = \frac{F(x_k)}{\Sigma_{i \in \{S\}} F(x_i)}$$

**Environment Generation on our Grid Implementation**: Environments are generated using Perlin Noise. With this method, we create varying, yet realistically consistent, environment traits, such as: altitude, temperature, rainfall, humidity, wind, and ocean current.

**Environment Generation on our Graph Implementation**: We also implemented our simulation in a graph, where each node was equivalent to a cell in our grid, but with random neighbors. Our graph implementation was primarily used to run supplementary experiments and create additional plots.

**Technical Outline**
Grid Implementation
- World class: Contains the color grid off which we render, parameters for whether to have disasters, a control for the overall reproductive rate of the trees. We initialize our random world and all plants randomly to one of the seven species at a random location.
    - env_step: the function called at each timestep. Iterates through all the world cells and asks each cell to perform a fitness competition. Spreads seeds for successfully reproducing trees, re-renders the world, and prevents seeds that go into the ocean from sprouting into a tree. However, waterborne seeds can still survive and float to another landmass, albeit with a lower probability. Saves fitness, population, and convergence data for plotting.
    - Spread_seed: determines the (1) number of seeds to spread (drawn from a geometric distribution as a function of its survival probability) and (2) spreads the seeds to a new coordinate by a jointly gaussian distribution.
- Soil class: Each soil instance contains a tree, a list of seeds, and would handle the fitness competitions when multiple seeds and trees competed for the soil to sprout.
    - Fitness_competition: with some probability function, picks the next plant occupant of itself among the competitors: the seeds that arrive at the soil instance *at the previous timestep* plus the current tree occupant.
    - Tree_fitness: calculates the fitness of its tree occupant.
- Generator
    - Perlin noise-based world generator. The random numbers generated for each cell would map to one of the 7 environments based on the thresholds we chose.

- Experiments for running some hefty CPU-warming, patience-destroying experiments on 5 main criteria:
    - World dimensions and its impact on biodiversity and convergence
    - Mutation variance and its impact on convergence
    - Tree reproductive model: many low-quality offspring or one high-quality offspring and its impact on convergence
    - Frequency of disasters and its impact on convergence
    - Number of initial trees planted in the world and its impact on convergence

Graph Implementation
- World Class:
    - Similar to that of our Grid implementation, but with more methods for randomly generating a graph given the number of continents and nodes.
    - Contains functions to change continent attributes midway through an experiment (as a disaster simulation) and getting the average plant attributes in a given continent (for plotting results).
- Soil Class:
    - Env_step: Creates a fitness competition at the soil instance.
    - Fitness competition: Among the seeds that arrived *at the previous timestep* and the current occupant of the soil instance, probabilistically selects the occupant of the soil instance for the next timestep.
    - Tree_reproduction: On every timestep, surviving trees reproduce seeds to adjacent nodes, whose traits are drawn from a distribution with the tree's traits as mean, and mutation_var variance. With some very small probability p_shift, this distribution is randomly shifted left or right by shift_magnitude.
- Experiments for convergence in average plant fitness, convergence to unchanging continent traits, and convergence to changing continent traits.

Baseline Implementation
- Please see Appendix I.

## **Results**
### **I. Natural Selection**
Our first and primary experiment was to see if plants converged to the environment they are best suited to. We assess fitness by the fitness equation defined above. On a 10-node graph with 3 "continents" (strongly-connected components), we recorded the following average plant fitness over time (higher is better).

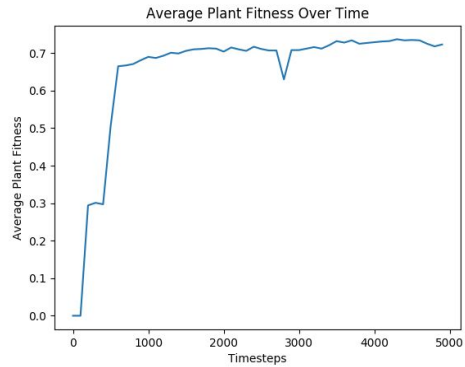

Figure 1: On a 3 continent graph, the average plant becomes better suited to its environment over time.

If we visualize the continent traits and evolving plant traits on the same graph, we can see the rate of convergence--how quickly the plant adapts to its continent traits.

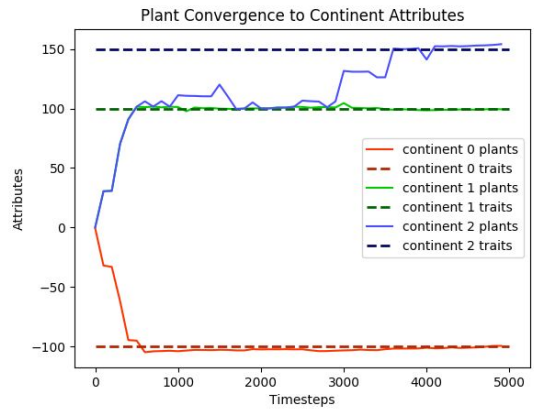

Figure 2: All plants were initialized to an attribute ~N(0,1) and converged to their respective environments which had attributes -100, 100, and 150 respectively. Note that continents 1 and 2's plants converged almost in lock-step, increasing at similar rates until around timestep 3000, when continent 2's plant traits diverged from continent 1's. This simulation was generated with mutation variance 0.05, shift_magnitude 10, and shift_probability 0.01.

On the grid visualization, we can see the plants converge to their environment traits. The more a plant is fit for its environment, the more similar its color will be to the "target image," which occurs in the case of perfect convergence:

| Start | Time = 10 | Time = 20 | Time = 40 | Time = 100 |

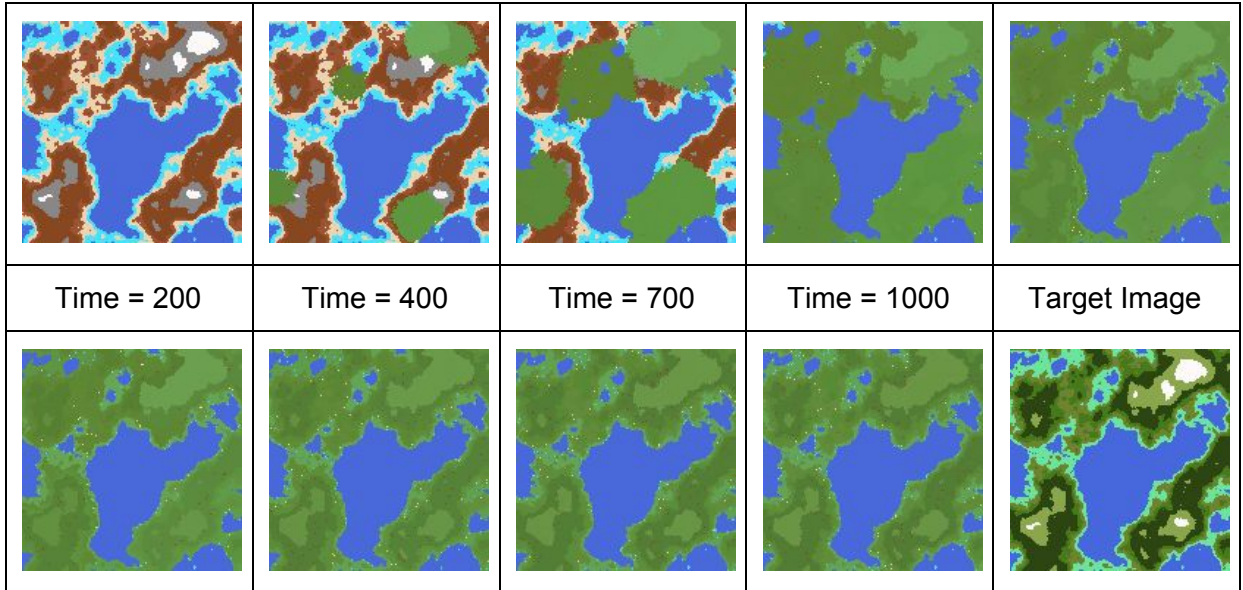

Figure 3: Convergence of plants traits to match environment as visualized in our grid world.

**Mutation Variance**

Holding all other parameters constant, we tested multiple values of mutation to see its effect on convergence of plant traits to the environment.

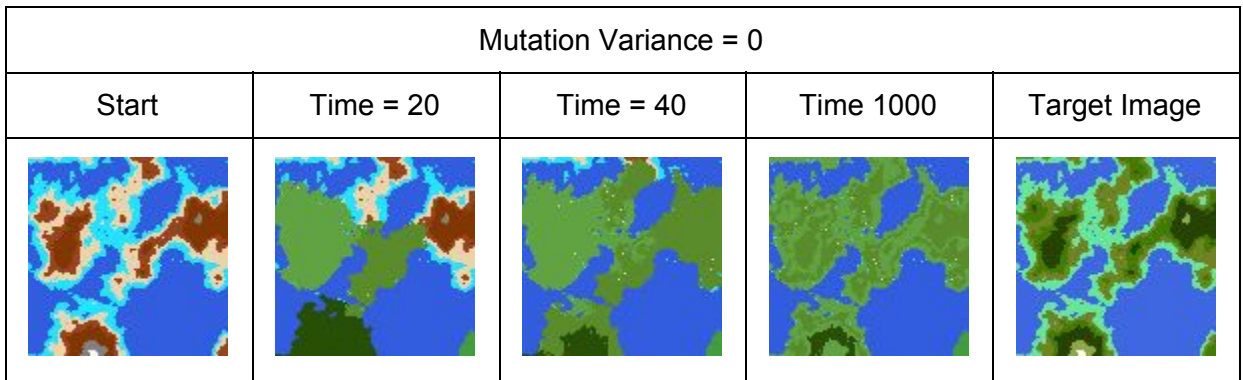

Figure 4: In this experiment, there were three distinct plant species that spread: dark green, traditional green, and light green. We can see that with no mutation variance (i.e. plant fitness cannot evolve), the good-fit plants drift to their own most-suitable environments.

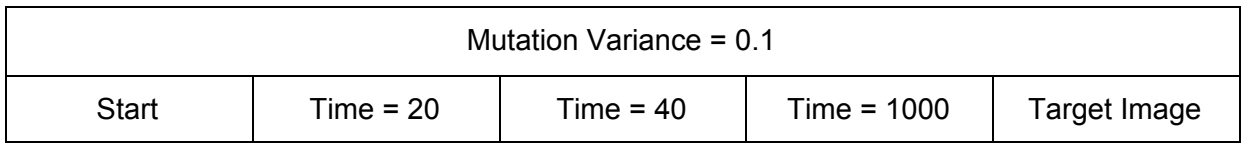

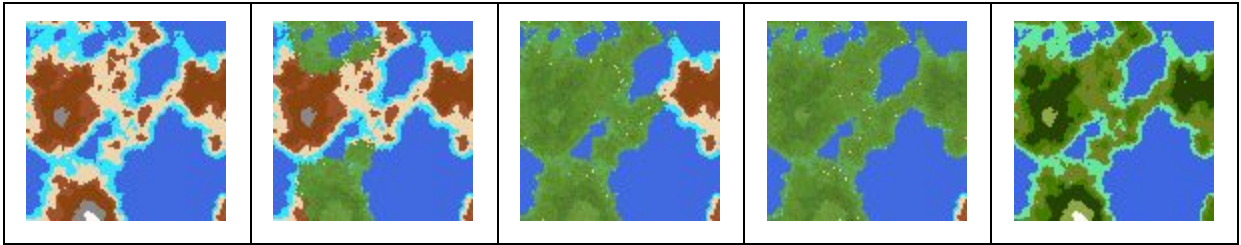

Figure 5: With an increased variance, plants begin to speciate. In addition to migrating to ideal habitats, the plants evolve to better fit their environments.

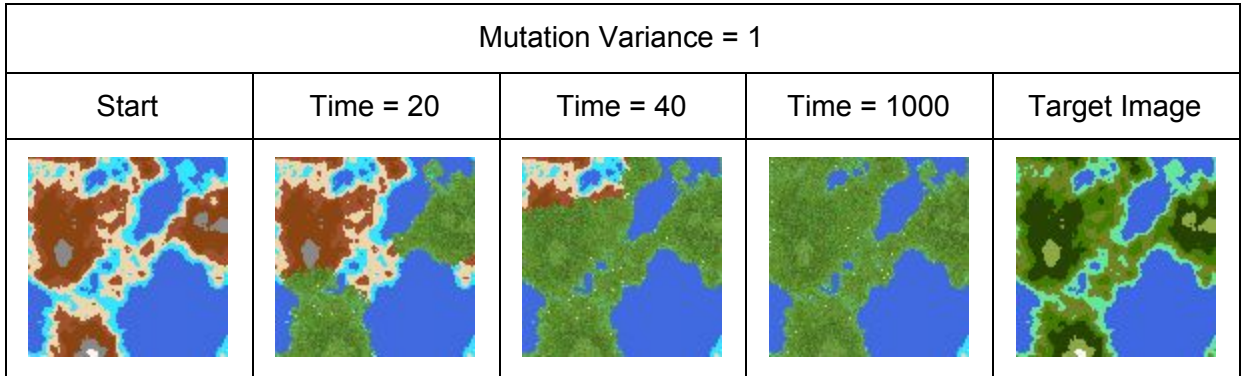

Figure 6: When the variance is high, the plant traits have a visibly large amount of variation, and converge to the traits of the continents at a much faster rate. This is observable in how the number and geographical distribution of species at time 40 and time 1000 is nearly identical.

Shown below is the convergence progress of plants given different mutation variances. Recall that there were 7 environments and 7 plant species, and each plant had a one-to-one mapping with its preferred environment. The definition of "convergence" we use is the proportion of plants in the entire world in their preferred environments:

$$Convergence = \frac{\text{Number Of Plants In Their Preferred Environment}}{\text{Total Numnber of Plants in the World}}$$

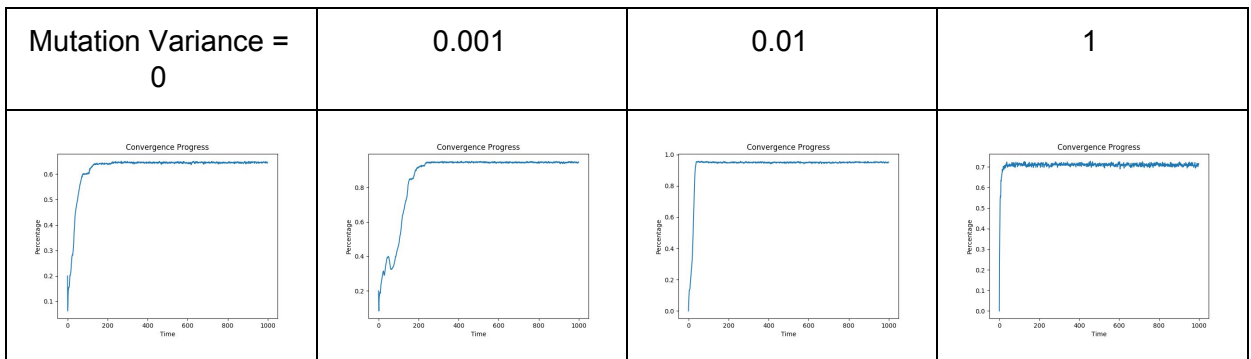

Figure 7: Convergence rates as mutation variance increases.

We found that without any evolution (no variation in offspring), plants will migrate to the areas for which they are best suited. When slight variation is added to offspring, we find that plant's trait converge to their environment. As potential variation in offspring increases, plants converge more quickly to environments for which they are best suited. However, an excessively high mutation variance can cause plants to not converge as closely to the continent traits as a lower mutation variance situation. As seen in the above table, a world with mutation variance of 0.01 converged much closer to 1 than a world with mutation variance of 1, in which the convergence plateaued at 0.7.

## II. r/K Selection

In this experiment, we compared K-selecting plants, which have one "quality" offspring (K-selection), and r-selecting plants, which have multiple "lower-quality" offspring.

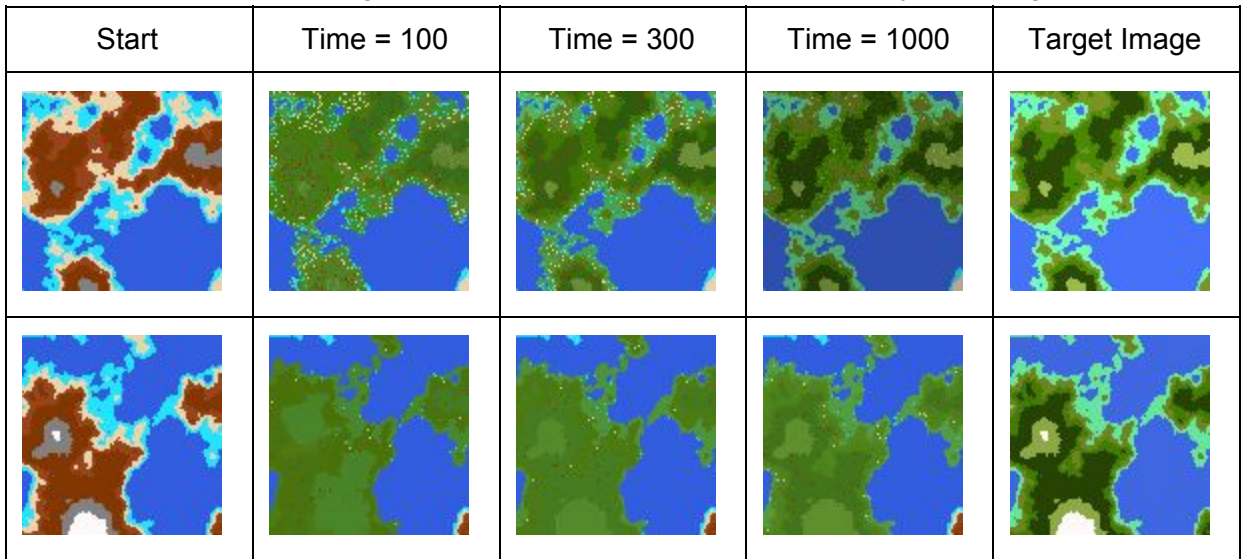

Figure 8: Top row: K-selection (few, quality offspring). Bottom: R-selection (many, low-quality offspring).

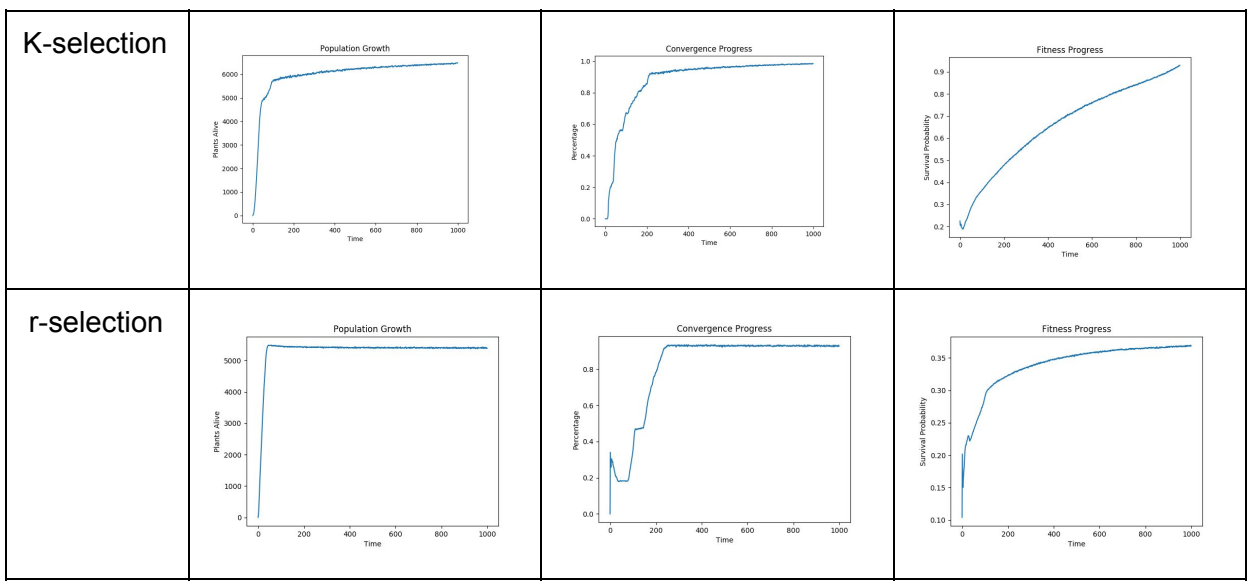

Figure 9: Change in population size, fitness and convergence rates for both r- and K-selection simulations.

In the K-selection simulation, population growth was slower, but, individual fitness growth climbed steadily throughout the length of the experiment. In the R-selection simulation, population plateaued very quickly, convergence was achieved slightly later than in K-selection, and fitness plateaued sooner.

### III. Disturbance Resilience

On our graph implementation, we simulated a disturbance by changing the continent attributes 40% and 80% of the way through the entire experiment and seeing if the plant would converge to the new traits. In most test trials, the plants converged.

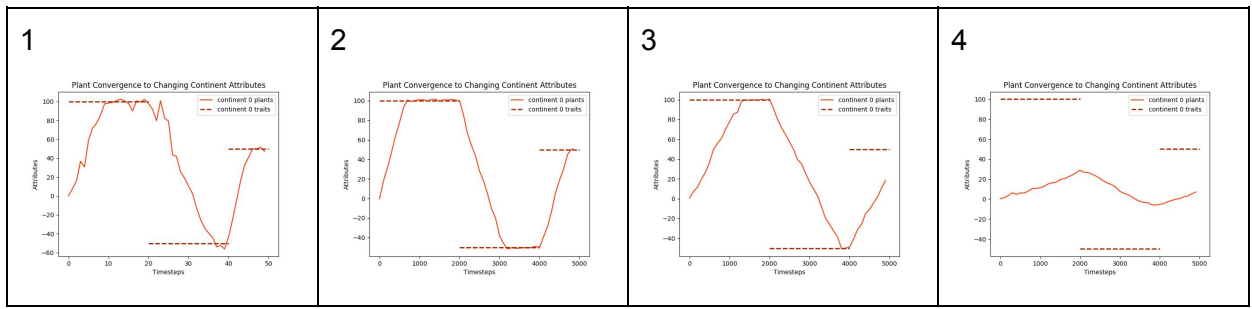

Figure 10:
1. Very high mutation variance in offspring, very low number of environment timesteps. (mutation_var = 15, total_steps = 50, shift_magnitude=0).
2. Mutation variance of 0.4 in offspring, 5000 environment timesteps. No shift_magnitude.
3. Mutation variance of 0.3. No shift magnitude. This variance is low enough such that the plants do not converge to the continent traits in time before the experiments end.
4. Mutation variance of 0.1. No shift magnitude. Plant traits are visibly affected by environment disturbances but do not converge to the continent traits on any occasion.

On our grid implementation, we change the continents to simulate a disturbance event. A similar pattern to the graph implementation is seen. A disturbance, such as flooding or earthquakes, occurs on every 200th timestep (bolded).

| Time = 100 | **Time = 200** | Time = 300 | **Time = 400** | Time = 500 |
|---|---|---|---|---|

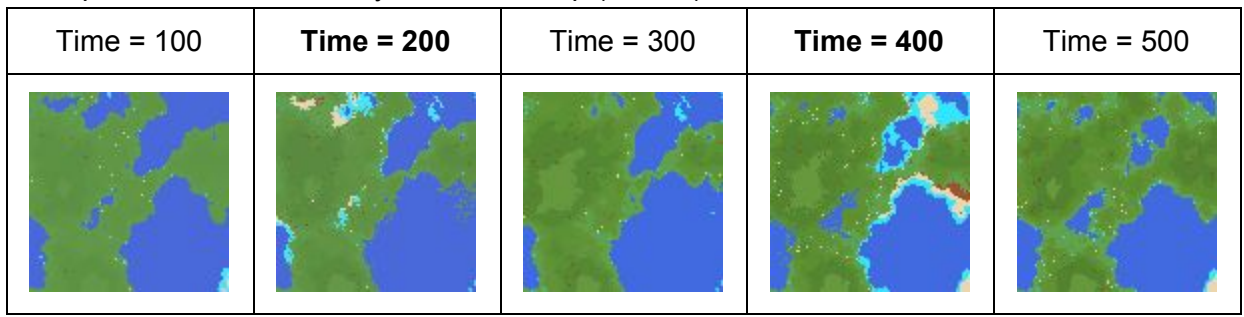

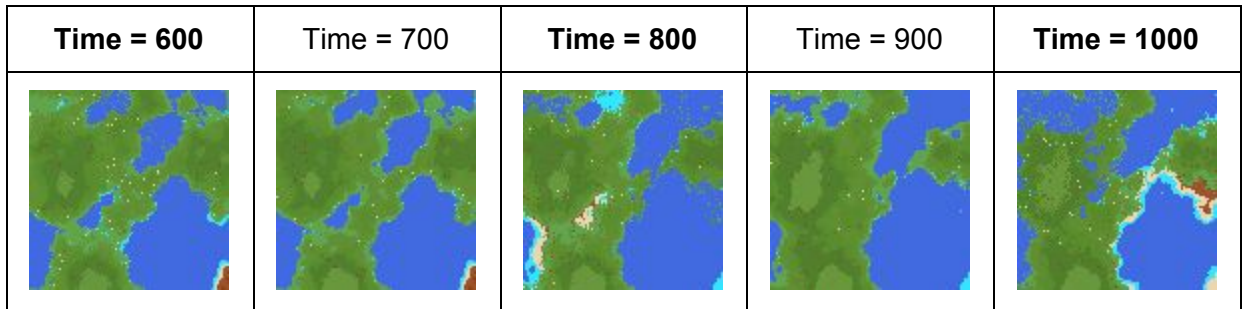

Figure 11: Grid world visualization of convergence with a disturbance event that occurs every 200 timesteps. Note: timesteps 200, 400, 600, 800 and 1000 are when the disturbance events occur, so it seems as if plants are growing in the water (they will likely die in the next few timesteps) following the disaster.

We tested whether the frequency of disturbances affected the plant's ability to converge. In all cases, plants rebounded to their pre-disturbance peak.

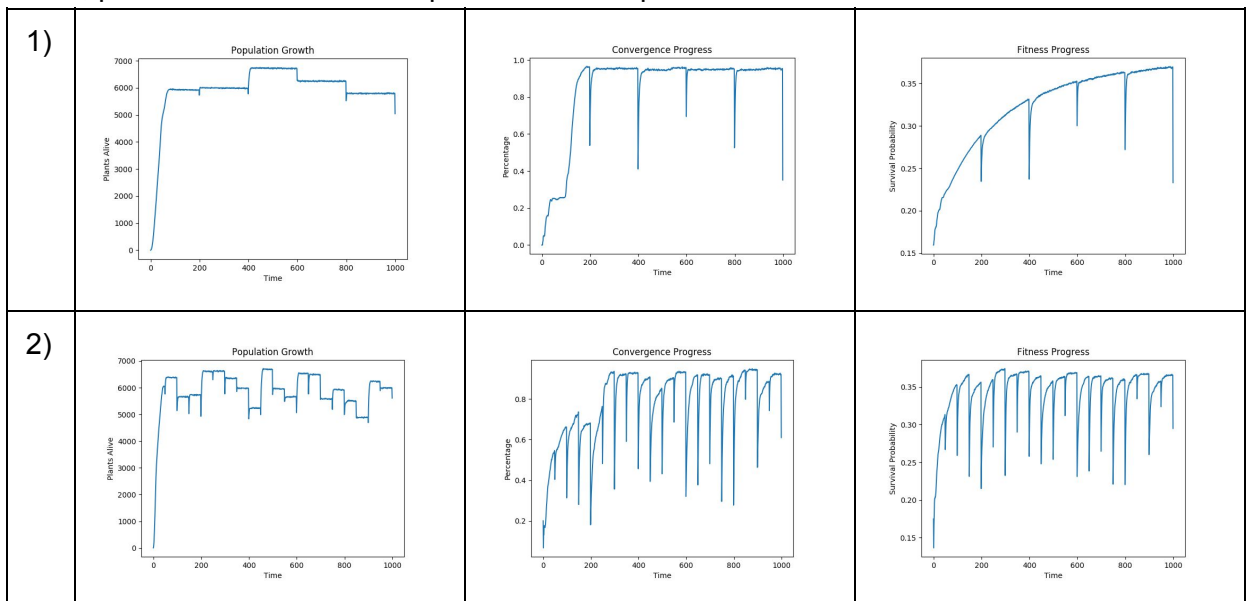

Figure 12: Population of plants in two disturbance event simulations. Row 1 has a disturbance event frequency of 200 timesteps. Row 2 has a disturbance event frequency of 50 timesteps.

In both the grid and graph implementation of our simulation, plants with a non-zero mutation variance parameters were able to adapt to disturbance events. Interestingly, as plant's fitness increases, the time it takes to recover from a disturbance event decreases, even if the disturbance event is more damaging.

**Technical challenges and resolutions**
1. **Achieving Convergence**. This was perhaps the hardest task on both the grid and graph implementations. There were many parameters we had to tune before we achieved convergence almost always. First, we found that the choice of fitness function was extremely important. We were originally using e^-x, but e^-0.25x worked better for aforementioned reasons, as well as being the probabilistic function for choosing the winner amongst many different seeds and plants at a square. In addition, the mutation variance was a constant knob we had to tune for speed to convergence--an extremely low mutation variance would take a long time to converge. Finally, on the graph implementation, if there was a continent "A" with only two nodes, and each of the two nodes were connected to other continents "B" and "C", it was extremely difficult for the plants on the continent "A" to converge because they were constantly being flooded by seeds, adapted to a different environment, from continents "B" and "C", which were larger and spread more seeds. How we dealt with this issue on the graph implementation was by adding p_shift and shift_magnitude variables, as mentioned briefly earlier in the paper. With some small probability p_shift, a seed's offspring's mean would be drawn from a gaussian not centered on its parent's traits, but on its parent's traits plus shift_magnitude, which would have randomly been either positive or negative. This simulated extremely rare, large-scale genetic drift and mutation events uncapturable by our standard mutation_variance parameter. After implementing p_shift and p_magnitude, graphs with even the difficult-to-converge setting of two-continent nodes were able to converge almost always.

2. **Rendering a randomized world**. Random number generators are extremely volatile and do not map well to smooth contours when creating a random 2D map. Thus, we looked into smoothed random noise libraries, especially Perlin and OpenSimplex. We spent a substantial amount of our time understanding the python noise library that we used as our Perlin random number generator to decide which of the seven categories each cell in our grid would be assigned to for smooth contours. The noise library does not provide a seed function, so for the same sized grids, every time it is initialized, it produces the same world. We overcame this by leveraging another noise library, Open Simplex, which also generated smooth contours. However, we preferred the random results of the Perlin world much more simply because they looked much better and more world-like. How we reconciled this was by generating two (D x D) sized grids with the Open Simplex library and one (D x D) grid with Perlin noise. Our resulting world was a simple weighted sum:

$$X, Y \sim OpenSimplex(D \times D)$$
$$Z \sim Perlin(D \times D)$$
$$InitializedWorld = c \cdot (X - Y) + Z, \text{where } c \ll 1$$

3. **Runtime**. We ran hundreds of experiments over the course of this project. Our final twenty experiments, which gave us many of the plots in this paper, took about 8 hours to run in total. Thus, a significant limitation of our project is runtime. After the initial few timesteps, the number of seeds and plants to keep track of explodes, which drastically slows down the speed of our simulation. There are also huge speed differences even between a (100 x 100) world (our default for the grid implementation) versus a (125 x 125) world.

**Conclusion**

Using both our graph and grid implementations, we leveraged interacting particles in a randomly generated to validate existing ecological theory through carefully thought out experiments.

By showing plants migrate to ideal habitats without evolution, and with evolution, species converge to the number of environments, we further justified Darwin's natural selection theory. We demonstrated the tradeoffs of r-/K-selection proposed by MacArthur and Wilson: r-selected species have higher population growth rates, while K-selected species have higher individual fitness. While plants are generally known to be extremely resilient organisms; our model confirmed this empirically.

In addition to the theories tested in our report, our model is well suited to test another well known ecological theory also from MacArthur and Wilson's 1967 publication, Island Biogeography. This theory says that an isolated, small island should have fewer species than a larger, less isolated island. We could test this by creating a mainland with large and small islands far and close to the mainland, running our experiment, and comparing the number of species of each island, in convergence, to its size and isolation.

Our simulation's use is not limited to the experiments conducted in this report. It has the potential to evaluate the diversity of future plant communities in the face of the changing climate. It could be used to predict the amount of time needed for a community to recover from a natural disaster such as a wildfire or earthquake. It could even be useful for evaluating human-plant interaction, if human activity were to be added as frequent, small scale disturbances.

## Appendix I: Baseline Project

| | | Complete Graph (n) | Ring Graph (n) |
|---|---|---|---|
| # Starting Diseases | n | 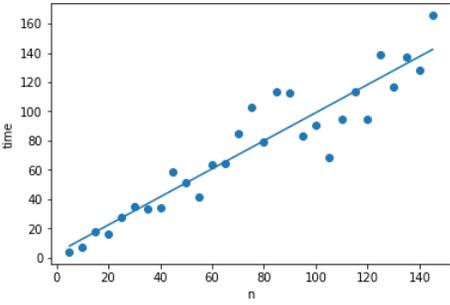<br>y = 0.96x + 3.1 | 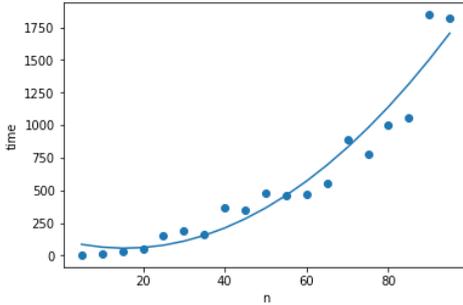<br>y = 0.26x² + -8.2x + 120.3 |
| | 2 | 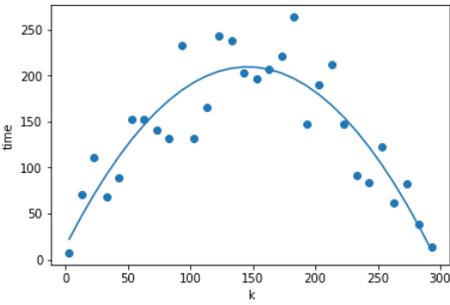<br>y = -0.009x² + 2.7x + 14.3 | 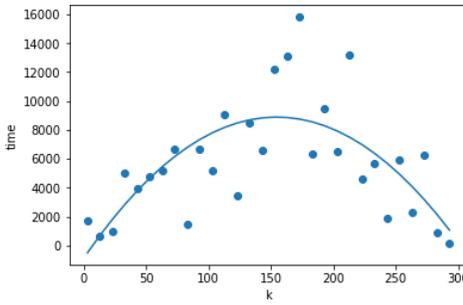<br>y = -0.4x² + 126.4x + -885.1 |

**Summary of Procedure and Results**

The aim of this experiment is to explore the relation between initial disease distribution and time it takes for one disease to dominate. For the first part, we initialized each human (node) to have its own disease. Connections between humans are modeled as (a) a complete graph and (b) a ring graph, both with n nodes. We tested graph sizes from 5 to 150 with a step of 5. We plotted the average of 8 trials on the graph.

With n starting diseases, on a complete graph, the graph size is linearly proportional to the one-disease takeover time. For a ring graph, the relationship was quadratic.

For the second part, we initialize the nodes to have one of two diseases: (a) a complete graph and (b) a ring, both with n nodes. The parameters of this environment are the number of people that begin with disease 1, denoted k on the graph. (This constraints the number of people that have disease 2 to be n - k). The size of graph (n) is fixed to be a big constant. We plot the average of 15 trials for each point on the graph. We ran the experiment for 99 values of k:

$$k \in \{\tfrac{n}{100}, ..., 99 * \tfrac{n}{100}\}, n = 300$$

The graph is symmetric, as we expect, with respect to k = 150. There is a quadratic relationship between k and one-disease-overtaking time, somewhat reminiscent of the entropy curve of Bernoulli variables with probability p, where the "disorder" is highest when the proportion of people affected by each disease is split 50/50.

On a ring graph with two starting diseases, the shape of the curve is roughly the same as the case with a complete graph, but scaled by roughly a factor of 100. For instance, when k = 150, the time for one disease to take over is more than 10000 compared to less than 500 on a similar-sized complete graph.

One possible explanation for this is that a ring has non-uniform distance between its nodes, unlike a complete graph where the distance between any two nodes is 1. Thus a ring graph exhibits "local" behaviors where people of the same disease are close to each other. Therefore the probability that the distribution of a disease changing is much lower. Consider the extreme case when half of the circle is of disease one and half is of disease zero. There are only 2 out of n edges that can change the disease distribution. Spreading diseases along all remaining edges does not actually pass a disease of one type to the other.

The complete graph, two initial diseases problem could be modeled easily as a birth-death CTMC.
- The states of our markov chain are the number of people with disease 1
- The two terminal states would be 0 and N, which means disease 0 or 1 takes over, respectively.
- For node k, we transition to node

$$\begin{cases} k+1, & \text{w.p. } \frac{k(n-k)}{n(n-1)} \\ k-1, & \text{w.p. } \frac{k(n-k)}{n(n-1)} \\ k, & \text{w.p. } 1 - \frac{2k(n-k)}{n(n-1)} \end{cases}$$

However, solving for an analytical solution is challenging.

**Implementation Details**
We implemented two versions of the simulation. For our first, we pre-generated exponentially distributed times and placed them in a list. This had extremely slow runtime (over an hour per trial), and we were unable to get a meaningful plot. For the second implementation, we used a min-heap to keep track of the minimum exponential clock time. We also maintained a list of the number of people with each disease, which sped up computation of number of diseases existing on the nodes. To stay true to the spirit of the simulation, we did not take shortcuts by utilizing the memoryless property of exponential random variables, though leveraging it might have greatly improved our runtime.